\documentstyle[12pt]{article}

\font\tenrm=cmr10
\font\tenit=cmti10
\font\elevenbf=cmbx10 scaled\magstep 1
 1
\font\elevenit=cmti10 scaled\magstep 1

\newcommand{\eps}{\varepsilon}
\newcommand{\barr}[1]{\overline{#1}}
\newcommand{\non}{\nonumber}
\newcommand{\Li}{\mbox{\rm Li}}
\begin{document}
\begin{titlepage}
\pagestyle{empty}
\renewcommand{\textfraction}{0}
\renewcommand{\topfraction}{1}
\renewcommand{\bottomfraction}{1}
\begin{flushright}
{\elevenbf TTP93-20\\
July 1993}
\end{flushright}
\begin{center}
\vglue 0.6cm
{ {\elevenbf        \vglue 10pt
ENERGY OF W DISTRIBUTION IN TOP QUARK DECAYS\footnote{Work partly
supported by the Polish
Committee for Scientific
Research (KBN) Grants 203809101 and 223729102, and
by EEC Contract ERB-CIPA-CT-92-2077}
\\}
\vglue 1.0cm
{\tenrm Marek JE\.ZABEK\\}
\baselineskip=13pt
{\tenit Institute of Nuclear Physics, Kawiory 26a, PL-30055 Cracow,
Poland\\}
\baselineskip=12pt
{\tenrm and\\}
\baselineskip=12pt
{\tenit Institut f. Theoretische Teilchenphysik, Universit\"at
Karlsruhe\\}
{\tenit D-76128 Karlsruhe, Germany\\}

\vglue 0.3cm
{\tenrm and\\}}
\vglue 0.3cm
{\tenrm Christoph J\"UNGER\\}
{\tenit Institut f. Theoretische Teilchenphysik, Universit\"at
Karlsruhe\\}
{\tenit D-76128 Karlsruhe, Germany\\}
\vglue 0.8cm

\end{center}

\vglue 0.3cm
\abstract{
A relatively simple analytical formula is derived for the energy
spectrum of $W$ boson in top quark decays $t\to Wb$ including
${\cal O}(\alpha_s)$ radiative corrections. We discuss the
accuracy of this formula and compare it to a more general
albeit more complicated one derived in \cite{CJK}~. A Monte
Carlo algorithm for generation of $W$ energy spectrum is briefly
described.}
\vglue 2.6cm
\begin{center}
{\tenit (Submitted for publication to Acta Physica Polonica B)}
\end{center}
\end{titlepage}

\pagestyle{normal}
{\large\bf\noindent 1. Introduction and summary}
\vglue 0.8cm
\baselineskip=14pt
\rm
Radiative QCD corrections to the energy distribution of
$\bar ff^\prime$
in $t\to b\bar ff^\prime $ and $t\to bg\bar ff^\prime$
decays have been calculated some time ago \cite{CJK}~.
In the meantime the lower limit for top quark mass $m_t$
has been pushed up by CDF and D0 collaborations well above
the threshold for $t\to bW$ channel. Although the results
of \cite{CJK} are applicable also in this regime it seems
reasonable and useful to derive a new formula assuming
dominance of decays into real $W$~. Such a formula, albeit
less general than that given in \cite{CJK}~, can be very
useful in studies of top quark physics at future $e^+e^-$
colliders \cite{Workshop}~.\par
\vglue 0.2cm

Our present approach is based
on an appropriately modified narrow width ($\Gamma_{_W}=0$)
approximation, where $\Gamma_{_W}$ is the width of $W$ boson.
In contrast to \cite{CJK}~, where the rates are manifestly
infrared finite, we introduce an explicit infrared cutoff
$\lambda$ on the effective mass squared of the system
(b quark + gluon)~. Thus the formula for
${\cal O}(\alpha_s)$ contribution from virtual gluons also depends on
$\lambda$. This apparent failure turns out to be an advantage
in Monte Carlo simulations, which are indispensable for a precise
determination of the strong coupling constant $\alpha_s$ and $m_t$
from the total $\sigma(e^+e^-\to\bar tt)$ and differential
cross sections \cite{SP,Sumino,JKT,Martinez}~. We wrote a Monte
Carlo program based on the results of the present article and
found a very good agreement with the formulae given in \cite{CJK}~.
\vglue 0.2cm

\par
We have checked also that massless $b$ approximation which is
known to be a satisfactory one for the total decay rate and
$m_t$ above 120 GeV \cite{JK1,Topw}~, cannot be used in calculations
of $W$ energy distribution. One reason is purely kinematical:
Born distribution in the narrow width approximation has a Dirac
delta shape, i.e. the energy of $W$ is fixed by two body
kinematics. A shift from realistic $m_b$~=~4.7~GeV to $m_b$~=~0
results in a non-negligible shift in this energy. We attempted
to correct the massless formula (see Appendix A) for this
trivial effect but the result was still a rather poor approximation.
Thus, we conclude that for realistic top and bottom masses one
has to use the complete ${\cal O}(\alpha_s)$
result rather than its massless approximation.
\vglue 0.2cm

\par
Our article is organized as follows.
In sect.2 we derive our formula for the enegy of $W$ distribution
in the narrow width approximation. In sect.3 we include a non-zero
$W$ width and describe our Monte Carlo program based on our
calculations. Then, we compare the results of this program with those
of \cite{CJK}~. In Appendix A our formulae for $m_b \rightarrow 0$ are given.

\pagebreak[4]

{\large\bf\noindent 2. Energy of W distribution in narrow width
approximation}
\vglue 0.6cm
We use throughout the same notation as in \cite{JK2}~. We stay in the
top quark rest frame and some variables are expressed in units of
$m_t$. In particular $y$ -- effective mass squared of $\bar ff^\prime$
system,  $z$ - the mass squared of $bg$ system and
$\eps= m_b/m_t$~. We define also the energy of $W$~:
\begin{equation}
x_{_W} = {E_{_W}\over m_t} = {1\over2} (1+y-z)
\label{xw}
\end{equation}
which for twobody Born decay mode $t\to bW$ is replaced by
\begin{equation}
\overline{x}_{_W} = {E_{_W}\over m_t} = {1\over2} (1+y-\eps^2)
\label{xwbar}
\end{equation}
and $x_{_W} \le \barr{x}_{_W}$. In general `barred' quantities are evaluated
at $z=\eps^2$. Other useful variables are:
\begin{equation}
w_3 = \sqrt{x_{_W}^2 - y} = {1\over2} \sqrt{\lambda(1,y,z)}
\label{w3}
\end{equation}
\begin{equation}
\lambda(u,v,w)=u^2+v^2+w^2-2(uv+uw+vw)
\end{equation}
the lenght of $W$ momentum\footnote{In unpolarized case discussed here
we choose $z$-axis in the direction of $W$.},
which is equal to
\begin{equation}
p_3 = \sqrt{x_h^2 - z}
\label{p3}
\end{equation}
the momentum of $bg$ system of energy $x_h=1-x_{_W}$,
the light cone variables
\begin{eqnarray}
w_\pm &=& x_{_W} \pm w_3  \nonumber \\
p_\pm &=& x_h \pm p_3
\label{pwpm}
\end{eqnarray}
and rapidities
\begin{eqnarray}
Y_{_W} &=& {1\over2}\ln{w_+\over w_-}   \nonumber \\
Y_p &=& {1\over2}\ln{p_+\over p_-}
\label{rap}
\end{eqnarray}

In the narrow width approximation $\gamma = \Gamma_{_W}/m_{_W}\to 0$
one can replace a factor resulting from $W$ propagator by Dirac delta function
\begin{equation}
{1 \over\pi}{\gamma y_0\over(y-y_0)^2+\gamma^2 y_0^2}
\longleftrightarrow
\delta(y-y_0)
\label{delta}
\end{equation}
where $y_0 = (m_{_W}/m_t)^2$~. Thus, when the decay through real $W$
dominates the effective mass is close to $\sqrt{y_0}$~.
In this section we consider $y$ as fixed and give the formula
for the differential rate
$\left.{{\rm d}\,\Gamma\over{\rm d}\,x_{_W}}\right|_y$~.
Then, in sect.3 we relax this assumption using (\ref{delta})~.
For fixed $y$ the energy distribution of $W$ including
${\cal O}(\alpha_s)$ corrections can be written as follows:

\begin{eqnarray}
\left.{{\rm d}\,\Gamma\over{\rm d}\,x_{_W}}\right|_y
&=&\frac{{\rm G}_F m_t^3}{8\sqrt{2}\pi}\left[\delta(z-\eps^2)
    \left({\rm F}_0-\frac{2\alpha_s}{3\pi}\tilde{{\cal G}}_1(y,\lambda)
       \right) \right. \non \\
   & & \left. \hspace{1.5cm} +\frac{2\alpha_s}{3\pi}
   \Theta(z-\eps^2-\lambda)
       \Theta(x_{_W}-\sqrt{y})\,\tilde{g}_1(y,z)\right]
\label{dGdx|y}
\end{eqnarray}
where $z$ is fixed through (\ref{xw})~,
\begin{eqnarray}
{\rm F}_0(y,\eps) &=&
{1\over 2}\sqrt{\lambda(1,y,\eps^2)}\,{\cal C}_0(y,\eps) \label{F0} \\
{\cal C}_0(y,\eps) &=&
4[(1-\eps^2)^2+y(1+\eps^2)-2y^2] \label{C0}
\end{eqnarray}
\begin{eqnarray}
\tilde{{\cal G}}_1
&=&g_1 {\cal C}_0(y,\eps) \barr{x}_h
+ g_2 {\cal C}_0(y,\eps) \barr{p}_3 +
             g_3 \barr{x}_h \barr{p}_3 + g_4 \barr{p}_3 +
             g_5 \barr{Y}_p + g_6    \non    \\
g_1&=&
            4{\barr{Y}_p}^2
            - 2\Li_2(\barr{w}_0)+4\Li_2\left({\barr{w}_-}/
                  {\barr{w}_+}\right) \non \\
& & -4\Li_2\left(\frac{\barr{p}_-
\barr{w}_-}{\barr{p}_+\barr{w}_+}
    \right)+2\Li_2(\barr{w}_+)-8\barr{Y}_p\barr{Y}_w \non
                           \\
& & +4\barr{Y}_p\left(2\ln\eps-\ln\lambda+2\ln\left(1-
                 {\barr{p}_-}/{\barr{p}_+}\right)-\ln y\right)
               \non  \\
g_2&=&4-6\ln\eps+4\ln\lambda \non \\
g_3&=&24(1-\eps^2)\ln\eps \non \\
g_4&=&8\,\left( -1 + 2\,{{\eps}^2} -
             {{ \eps}^4} - y - {{\eps}^2}\,y +
             2\,{y^2} \right)  \non \\
& &  -\Big[ 18\,\eps^2\,\left(2y^2-y-1\right)+
          4\,\eps^2{\cal C}_0(y,\eps) \Big]\,\frac{\lambda}{
                \eps^2\left(\eps^2-\lambda\right)} \non \\
g_5&=&4\,\left( -1 + {{\eps}^2} +
           {{\eps}^4} - {{\eps}^6} - 3\,y +
           2\,{{\eps}^2}\,y - 3\,{{\eps}^4}\,y +
           9\,{y^2} +  9\,{{\eps}^2}\,{y^2} -
           5\,{y^3} \right) \non  \\
g_6&=&\Big[9\,\eps^2\left(2y^2-y-1\right)+2\,
           \eps^2{\cal C}_0(y,\eps)\Big]\,\frac{1+y}{1-y}\,
           \ln\left(1+{\lambda}/{\eps^2}\right)- \non \\
& & \Big[7\,\left(2y^2-y-1\right)
      +2\,{\cal C}_0(y,\eps)\Big]\,
     \left(1-y\right)\,\ln\left(1+{\lambda}/{\eps^2}\right)
\label{Gtilde}
\end{eqnarray}
and
\pagebreak[4]
\begin{eqnarray}
\tilde{g}_1&=& 2\,a_1\,p_3(z) + 4\,a_2\,Y_p(z) -
         \frac{4\,\eps^4 {\cal C}_0(y,\eps)}{z^2(z-\eps^2)}\,p_3(z)
         + \frac{4\,\barr{x}_h {\cal C}_0(y,\eps)}{z-\eps^2}\,Y_p(z)
                     \non \\
     a_1&=& \frac{\eps^2}{z^2} \,\left[-9 + 15\,\eps^2
                  - 8\,\eps^4 - y\,(9+7\,\eps^2-18y) \right]
                \non  \\
            && +\, \frac{1}{z} \,\left[-7 + \eps^2\,(20-
                            5\eps^2-11y) + 7y\,(2y-1) \right]  \non \\
            && +\, 2y-3(1+\eps^2) \non \\
     a_2&=& 2 + \eps^2(\eps^2-5) - 2y\,(2y-1) +
               \left(1 + \eps^2 +2y \right)\,z
\end{eqnarray}

\noindent
In the above formulae $\lambda$ denotes an infrared cutoff on the effective
mass $z \ge \eps^2+\lambda$. Let us sketch now the derivation of eq.
(\ref{dGdx|y}): The contribution resulting from real gluon radiation
($\Theta$-piece) is obtained by direct integration of the fully differential
decay rate, whereas the (Born + soft) contribution ($\delta$-piece) is
derived from the requirement that integrating (\ref{dGdx|y})
over $x_{_W}$ one obtains the narrow width limit of the expression for the
total rate given in \cite{JK1}; see also (\ref{dGdy}) in the following
section. The formula (\ref{dGdx|y}) simplifies considerably in the limit
$m_b \rightarrow 0$, c.~f. Appendix A. However, for realistic b quark masses
this is a rather poor approximation.
\vglue 0.6cm
{\large\bf\noindent 3. Finite W width and results}
\vglue 0.6cm
We generalize now the results of the previous section and include
a nonzero $W$ width. Let us note that $y$ is not fixed for $\gamma\ne0$.
The double differential distribution
\begin{equation}
{{{\rm d}\,\Gamma}\over{{\rm d}\,x_{_W}\,{\rm d}\,y}}
= {\gamma \over\pi}{y_0\over(y-y_0)^2+\gamma^2 y_0^2}
{{\rm d}\,\Gamma\over{\rm d}\,x_{_W}}\Bigg|_y
\label{dGdxy}
\end{equation}
is, however, closely related to the narrow width result (\ref{dGdx|y}), c.~f.
(\ref{delta}). We use the standard model result for $\Gamma_{_W}$:
\begin{equation}
\Gamma_{_W}= {{\rm G}_F {m_{_W}}^3\over 6\sqrt{2}\pi}
\left( 9 + 6 {\alpha_s\over\pi}\right)
\label{GammaW}
\end{equation}
Integrating (\ref{dGdxy}) over $x_{_W}$ we obtain of course
the standard model result for
${\rm d}\,\Gamma/{\rm d}\,y$ \cite{JK1,Topw} \footnote{we derived
(\ref{dGdx|y}) using this condition.}
\begin{equation}
{{\rm d}\,\Gamma\over{\rm d}\,y} =
{{{\rm G}_F}^2 {m_t}^5\over 192\pi^3}
\left( 9 + 6{\alpha_s\over\pi}\right)
 {1\over (1-y/y_0)^2+\gamma^2}
\left[ {\rm F}_0(y,\eps) - {2\alpha_s\over 3\pi}
{\rm F}_1(y,\eps)
\right]
\label{dGdy}
\end{equation}
$0 \le y \le (1-\eps)^2$~, where $F_0$ is defined in (\ref{F0}) and
\begin{eqnarray}
{\rm F}_1(y,\eps)&=&
          \frac{1}{2}{\cal C}_0(y,\eps)(1+\eps^2-y)
          \Big[ 2\pi^2/3 +4{\rm Li_2}\,(u_w)
         -4{\rm Li_2}\,({u_q})
          \non \\
& &       \hspace{1cm} -4{\rm Li_2}\,({u_q}{u_w})
        -4\ln{u_q}\ln(1-{u_q})
                \non \\
& &       \hspace{1cm} -2\ln{u_w}\ln{u_q}+\ln{y}\ln{u_q}
          +2\ln\eps\ln{u_w}  \Big]
          \non \\
& &       -2{\rm F}_0(y,\eps)
         \left[      \ln{y}+3\ln\eps-2\ln\lambda(1,y,\eps^2) \right]
               \non \\
& &       +4(1-\eps^2)\left[ (1-\eps^2)^2
           +y(1+\eps^2)-4y^2 \right]  \ln{u_w}
             \non \\
& &     +\left[ 3-\eps^2+11\eps^4-\eps^6+
         y(6-12\eps^2+2\eps^4) - \right.
                   \non \\
& &      \left. \hspace{1cm} y^2(21+5\eps^2)+12y^3 \right]\ln{u_q}
               \non \\
& &     + 6\sqrt{\lambda(1,y,\eps^2)}
           (1-\eps^2)(1+\eps^2-y)\ln\eps
                           \non \\
& &     +\sqrt{\lambda(1,y,\eps^2)}\left[ -5+22\eps^2
     -5\eps^4- 9y(1+\eps^2)+6y^2\right]
\label{F1}
\end{eqnarray}
where
\begin{equation}
{u_q}=\frac{\barr{p}_-}{\barr{p}_+}\hspace{1cm}
{u_w}=\frac{\barr{w}_-}{\barr{w}_+}
\end{equation}

\begin{figure}[h]
\begin{center}
\mbox{\epsfig{file=paper1.eps,width=12.cm,height=4.5cm,bbllx=0.cm,bblly=1.5cm,bb
\caption{Normalized energy distribution of $ W $: $ {\Gamma}^{-1}\,
     {\rm d}\Gamma/{\rm d}x_{_W} $ for $ m_t=120 $~GeV evaluated from
      (\protect\ref{dGdxnew})}
\end{center}
\end{figure}

We can also integrate (\ref{dGdxy}) over $y$. In this way we obtain a new
formula for
\begin{equation}
\frac{{\rm d}\Gamma}{{\rm d}x_{_W}}=\int\limits_0^{(1-\eps)^2}{\rm d}y\;
      \frac{{\rm d}\Gamma}{{\rm d}x_{_W}\;{\rm d}y}
\label{dGdxnew}
\end{equation}
We compared (\ref{dGdxnew}) with the result of \cite{CJK} and found a perfect
numerical agreement for small $\lambda$. For $\lambda=10^{-8}$ the relative
error is $10^{-6}$.

The formulae (\ref{dGdx|y}) and (\ref{dGdy}) can be also used as a starting
point for Monte Carlo simulations. A key observation is that (\ref{dGdy})
gives the distribution of $y$ for $0 \le y \le (1-\eps)^2$ whereas
(\ref{dGdx|y}) gives relative probabilities for $x_{_W}$ at fixed $y$.
The distribution (\ref{dGdxy}) can be generated as follows: $y$ is
generated first according to (\ref{dGdy}). Then, for given $y$, $x_{_W}$ is
generated according to (\ref{dGdx|y}). Both steps can be performed by a
standard combination of importance sampling and von Neumann rejection. The
only difficulty is to find the value of the infrared cutoff $\lambda$ such
that $\lambda$ is small enough and the $\delta$-piece in (\ref{dGdx|y})
remains positive. This well known difficulty limits relative accuracy of our
Monte Carlo to about 1\%.

In Fig. 1 we show the normalized distribution
${\Gamma}^{-1}\,{\rm d}\Gamma/{\rm d}x_{_W}$ for $m_t=120$~GeV
obtained from (\ref{dGdxnew}) for $\lambda=10^{-8}$. The curve obtained using
the result from \cite{CJK} is identical up to the resolution of the plot.
\begin{figure}
\begin{center}
\mbox{\epsfig{file=paper2.eps,width=12.cm,height=4.5cm,bbllx=0.cm,bblly=1.5cm,bb
\caption{Comparison of Monte Carlo for infrared cutoff
         $\lambda=3 \cdot 10^{-4}$
         and analytic result (\protect\ref{dGdxnew}) for $\lambda=10^{-8}$.
         The ratio
         (\protect\ref{ratio}) is shown as function of $x_{_W}$.}
\end{center}
\end{figure}

In Fig. 2 we compare the analytical result (\ref{dGdxnew})
($\lambda=10^{-8}$) with our Monte Carlo program for
$\lambda=3 \cdot 10^{-4}$.
We plot the ratio (in percent)
\begin{equation}
1-\frac{{\rm d}\Gamma/{\rm d}x_{_W}|_n}{{\rm d}\Gamma/{\rm d}x_{_W}|_a}
\label{ratio}
\end{equation}
where the subscript `n' refers to the Monte Carlo and `a' to the result
obtained from (\ref{dGdxnew}).
\newpage
\vglue 0.6cm
{\large\bf\noindent Appendix A}
\vglue 0.6cm
In this Appendix we give the limit of (\ref{dGdx|y}) for $m_b \rightarrow 0$.
These formulae can be used for $t \rightarrow s$ or $t \rightarrow u$
transitions. It would be, however, a rather poor approximation to use these
formulae for the dominant $t \rightarrow b$ transition. For
$\eps \rightarrow 0$ (\ref{dGdx|y}) is replaced by
\begin{eqnarray}
  \frac{{\rm d}\Gamma}{{\rm d}x_{_W}}\Bigg|_y&=&\frac{{\rm G}_F {m_t}^3}
        {8\sqrt{2} \,\pi}\,
  \left\{\delta(z)\,F_0 \left[1 - \frac{2\,\alpha_s}{3\,\pi} \,
           G_1(y,\lambda) \right] \right.\non \\
& &   \left. + \frac{2\,\alpha_s}{3\,\pi} \,\Theta(z-\lambda)\,
      \Theta \left(x_{_W}-\sqrt{y}\right)\,g(z,y)   \right\}
\end{eqnarray}
\begin{equation}
  F_0=2{\left(1-y\right)}^2\,\left(2y+1\right)
\end{equation}
\begin{eqnarray}
    G_1(y,\lambda)&=& \frac{2}{3}\,\pi^2 + \frac{5}{2} + 2\,\Li_2(y) +
                      4\,\ln^2(1-y) - 7\,\ln(1-y) + \ln^2(\lambda) \non  \\
                  & &  + \frac{1}{2}\,\ln(\lambda)\, \left[
                      7-8\,\ln(1-y) \right] + \frac{5+4y}{1+2y}\,\ln(1-y)
                              \non \\
                        \\
     g(z,y)&=& p_3(z)\left[2\,(2y-3) - 14\,(1-y)(2y+1)\,\frac{1}{z} \right]
                     \non \\
           & & + \,4\,Y_p(z)\left[(2y+1)(2-2y+z) + \frac{F_0}{z} \right]
\end{eqnarray}
Then, integrating the double differential distribution (\ref{dGdxy}) over $y$
one obtains ${\rm d}\Gamma/{\rm d}x_{_W}$ in the massless limit $m_b=0$.
\newpage
\vglue 0.4cm


\begin{thebibliography}{9}
\bibitem{CJK}
      A. Czarnecki, M. Je\.zabek and J.H. K\"uhn,
      {\elevenit Acta Phys. Polon.}
      {\elevenbf B20} (1989) 961;
      (E) {\elevenbf B23} (1992) 173
\bibitem{Workshop}
      P.M.~Zerwas (ed.), {\elevenit $e^+e^-$ Collisions at 500 GeV:
      The Physics Potential}, DESY Orange Report DESY 92--123A,
      Hamburg 1992
\bibitem{SP}
      J.M.~Strassler and M.E.~Peskin,
      {\elevenit Phys. Rev.} {\elevenbf D43} (1991) 1500
\bibitem{Sumino}
      Y.~Sumino, K.~Fujii, K.~Hagiwara, H.~Murayama and C.--K.~Ng,
      {\elevenit Phys. Rev.} {\elevenbf D47} (1992) 56
\bibitem{JKT}
      M.~Je\.zabek, J.H.~K\"uhn and T.~Teubner,
      {\elevenit Z. Phys. }{\bf C56} (1992) 653;
      M. Je\.zabek and T. Teubner,
      {\elevenit Z. Phys. }{\bf C} (1993) in print;
      Karlsruhe Univ. preprint TTP 93-11
\bibitem{Martinez}
      M. Martinez et al., in {\elevenit Proceedings of
      the Workshop on Physics
      and Experiments at Linear $e^+e^-$ Colliders,
      Waikoloa, Hawaii, April 1993}.
\bibitem{JK1}
      M. Je\.zabek and J.H. K\"uhn,
      {\elevenit Nucl. Phys.} {\elevenbf B314} {(1989)} 1.
\bibitem{Topw}
      M. Je\.zabek and J.H. K\"uhn,
      {\elevenit Phys. Rev.} {\elevenbf D}(Rapid Commun.) (1993)
      in print; Karlsruhe Univ. preprint TTP 93-4.
\bibitem{JK2}
      M. Je\.zabek and J.H. K\"uhn,
      {\elevenit Nucl. Phys.} {\elevenbf B320} {(1989)} 20.
\end{thebibliography}
\end{document}